\documentclass[12pt]{article}

\begin{document}
\begin{center}
{\bf EFFECTIVE LAGRANGIAN AND DYNAMICAL SYMMETRY BREAKING IN
$SU(2)\otimes U(1)$ NJL MODEL}\\
\vspace{5mm}
 S.I. Kruglov \\
\vspace{5mm}
\textit{International Educational Centre, 2727 Steeles Ave. W, \# 104, \\
Toronto, Ontario, Canada M3J 3G9}
\end{center}

\begin{abstract}
Dynamical symmetry breaking and the formation of scalar
condensates in the $SU(2)\otimes U(1)$ Nambu-Jona-Lasinio model
with two coupling constants has been studied in the framework of
the mean-field approximation. The bosonization procedures of the
model are performed using the functional integration method. The
possibility of the spontaneous CP symmetry breaking in the model
under consideration has been shown. The mass spectrum of the bound
states of fermions, as well as the effective Lagrangian of
interacting scalar and pseudoscalar mesons are obtained.

PACS numbers: 11.30.Qc; 12.39.Fe; 24.85.+p.
\end{abstract}

One of the modern problems is the investigation of nonperturbative
effects in quantum chromodynamics (QCD) and obtaining the
effective hadron Lagrangians. This area of low energy is difficult
to study in the framework of QCD because coupling constant
$\alpha_s$ is not a small parameter. Therefore, in this domain,
some phenomenological models are used. Among them are a model
based on local effective chiral Lagrangians (ECL) \cite{Weinberg},
\cite{Gasser}, \cite{Buchvostov}, a model of instanton vacuum
\cite{Shuryak}, \cite{Hooft} (see also \cite{Kruglov90}), the
Nambu-Jona-Lasinio (NJL) models \cite{Nambu} which use a contact
four-quark interactions, etc.. In this letter the possibility of
the spontaneous CP symmetry breaking in the NJL model is shown. It
should be noted that the electric dipole moment of particles
violates CP invariance and can be explained by the $\theta$-term
of QCD. Although the effect of CP violation in strong interactions
is small, the investigation of mechanisms of CP violation is
important.

Let us consider a four-fermion model with scalar-scalar and
pseudoscalar-pseudoscalar interactions and possessing the internal
symmetry group $SU(2)\otimes U(1)$ and two coupling constants:
\begin{equation}
{\cal L }(x)=- \overline{\psi}^n (x)(\gamma_\mu \partial_\mu
+m_0)\psi^n (x) +\frac{F}{2}\left[ \overline{\psi}^n(x)\psi^n
(x)\right]^2 - \frac{G}{2}\left[ \overline{\psi}^n (x)\gamma_5
\tau^a\psi^n (x)\right]^2 , \label{1}
\end{equation}
where $\tau^a$ ($a=1,2,3$) are the Pauli matrices, $\partial_\mu
=(\partial/\partial x_i ,-i\partial/\partial x_0)$ ($x_0$ is the
time), $m_0=\mbox{diag}(m_{01},m_{02})$; $m_{01}$, $m_{02}$ are
the current masses of fermions (quarks), $\gamma_\mu$ are the
Dirac matrices, $\gamma_5 =\gamma_1\gamma_2\gamma_3\gamma_4$,
$\psi$ is the doublet of fermions ($u$, $d$ quarks). Here
summation over colour fermion degrees of freedom $n=1,2,...,N_C$
has been performed. We also assume equal current masses of
fermions, $m_{01}=m_{02}\equiv m_0$. Lagrangian (1) is not
invariant under $\gamma_5$-chiral transformations, i.e. $U_A
(1)$-symmetry is broken. Note that the violation of
$U_A(1)$-symmetry allows one to solve the problem of the
differences of masses of $\eta$ and $\eta'$ mesons in the
framework of $SU(3)_f\bigotimes SU(3)_f$-model. We shell
investigate dynamical symmetry breaking (DSB) and the mass
formation using functional integration method \cite{Pervushin}.

Considering the generating functional for Green's functions
\begin{equation}
Z[ \overline{\eta},\eta]=N_0\int D \overline{\psi}D\psi
\exp\left\{i\int d^4 x\left[{\cal L}(x)+  \overline{\psi}^n
(x)\eta^n (x)+ \overline{\eta}^n (x)\psi^n (x)\right]\right\} ,
\label{2}
\end{equation}
where $ \overline{\eta}^n$, $\eta^n$ are external sources, and
redefining the constant $N_0$, one can represent Eq. (2) as
follows:
\[
Z[\overline{\eta},\eta]=N\int D \overline{\psi}D\psi D\Phi_0
D\tilde{\Phi}_a \exp\biggl\{i\int d^4 x\biggl[-\overline{\psi}^n
(x)\biggl[\gamma_\mu
\partial_\mu +m_0 -f_0\Phi_0 (x)
\]
\vspace{-7mm}
\begin{equation} \label{3}
\end{equation}
\vspace{-7mm}
\[
-ig_0\tilde{\Phi}_a (x)\gamma_5 \tau^a\biggr]\psi^n
(x)-\frac{\mu^2}{2}\tilde{\Phi}_a^2 (x) - \frac{M^2}{2}\Phi_0^2
(x) +\overline{\psi}^n (x)\eta^n (x) +\overline{\eta}^n (x)\psi^n
(x) \biggr]\biggr\} .
\]
 We define constants $F=f_0^2/M^2$, $G=g_0^2/\mu^2$, where
 $f_0$, $g_0$ are dimensionless coupling constants and constants $M$,
 $\mu$ are mass-dimensional. One can integrate the functional integral
 in Eq. (3) over the fermion fields $\overline{\psi}$, $\psi$ and obtain
\begin{equation}
Z[\overline{\eta},\eta]=N\int D\Phi_0 D\tilde{\Phi}_a
\exp\biggl\{iS[\Phi] + i\int d^4 xd^4 y \overline{\eta}^n (x)S_f
(x,y)\eta^n (y)\biggr\} , \label{4}
\end{equation}
\[
S[\Phi]=-\frac{1}{2}\int d^4 x\left[M^2 \Phi_0^2 (x)+\mu^2
\tilde{\Phi}_a^2 (x)\right]
\]
\vspace{-8mm}
\begin{equation} \label{5}
\end{equation}
\vspace{-8mm}
\[
-i \mbox{tr} \ln \left[1+\widehat{G}\left(f_0\Phi_0
(x)+ig_0\tilde{\Phi}_a (x)\gamma_5\tau^a\right) \right] ,
\]
where we introduce the effective action for bosonic collective
fields $S[\Phi]$. The fields $\Phi_0 (x)$, $\tilde{\Phi}_a (x)$
can be identified with the $\sigma$-meson and triplet of
pseudoscalar $\pi_a$-mesons \cite{Nambu}. The Green functions for
free fermions $\widehat{G}$, and for fermions in the external
collective fields $S_f (x,y)$ are solutions of the equations
\begin{equation}
(\gamma_\mu \partial_\mu +m_0)\widehat{G}(x,y)=-\delta(x-y) ,
 \label{6}
\end{equation}
\begin{equation}
\left[\gamma_\mu \partial_\mu + m_0 -f_0\Phi_0 (x)-
ig_0\tilde{\Phi}_a (x)\gamma_5\tau^a\right]S_f (x,y)=\delta(x-y) .
 \label{7}
\end{equation}
In four-fermion model under consideration, the symmetric vacuum is
not stable. As a result, the physical vacuum is reconstructed, and
the appearance of the condensates leads to DSB of the initial
$SU(2)\bigotimes U(1)$-symmetry. These condensates are
coordinate-independent fields $\Phi_0$, $\tilde{\Phi}_a$ (the mean
field approximation) which obey Eq. (7). Eq. (7) in the momentum
space for condensates $\Phi_0$, $\tilde{\Phi}_a$ becomes
\begin{equation}
(i\widehat{p} -A)S_f (p)=1 ,
 \label{8}
\end{equation}
where $\widehat{p}=p_\mu \gamma_\mu$, $p_\mu=(\textbf{p},ip_0)$,
$A=-m_0 +f_0\Phi_0+ig_0\tilde{\Phi}_a \gamma_5\tau^a$. Using the
gauge $\Phi_0\neq 0$, $\tilde{\Phi}_3\neq 0$,
$\tilde{\Phi}_1=\tilde{\Phi}_2=0$ in which the matrix $A$ is
diagonal, we obtain from Eq. (8) the Green function
\begin{equation}
S_f (p)= \mbox{diag}\left(
\frac{-i\widehat{p}+m_1+ig_0\tilde{\Phi}_3 \gamma_5}{p^2 +m^2},
\frac{-i\widehat{p}+m_1-ig_0\tilde{\Phi}_3 \gamma_5}{p^2 +m^2}
\right) ,
 \label{9}
\end{equation}
where the masses of fermions are defined as follows
\begin{equation}
m_1=m_0-f_0\Phi_0 ,\hspace{0.3in}m^2=m_1^2 +g_0^2 \tilde{\Phi}_3^2
.\label{10}
\end{equation}
If the current fermion masses $m_0=0$, the dynamical fermion
masses $m\neq 0$. It should be noted that the components
containing the term $ig_0\tilde{\Phi}_3 \gamma_5$ in Eq. (9)
violate CP parity.

To obtain the vacuum condensates $\Phi_0$, $\tilde{\Phi}_3$ from
Eq. (5), one finds the equations for the fields $\Phi_0 (x)$,
$\tilde{\Phi}_3 (x)$:
\[
\frac{\delta S[\Phi]}{\delta\Phi_0 (x)}=-M^2\Phi_0 (x)+if_0
\mbox{tr}S_f (x,x)=0 ,
\]
\vspace{-8mm}
\begin{equation} \label{11}
\end{equation}
\vspace{-8mm}
\[
\frac{\delta S[\Phi]}{\delta\tilde{\Phi}_a
(x)}=-\mu^2\tilde{\Phi}_a (x)- g_0 \mbox{tr}\left[S_f
(x,x)\gamma_5\tau^a\right]=0 .
\]
With the help of Eqs. (9), (11), we obtain gap equations for the
vacuum condensates:
\[
M^2\Phi_0 =2f_0 I m_1 ,
\]
\vspace{-8mm}
\begin{equation} \label{12}
\end{equation}
\vspace{-8mm}
\[
\mu^2\tilde{\Phi}_3 =-2g_0^2 I\tilde{\Phi}_3 ,
\]
where
\begin{equation}
I =\frac{iN_C}{4\pi^4}\int \frac{d^4 p}{p^2
+m^2}\hspace{0.5in}(d^4 p=id^3 pdp_0 ) . \label{13}
\end{equation}
The quadratically divergent integral in Eq. (13) can be evaluated
with help of the momentum cutoff $\Lambda$, and gap equations (12)
possess non-trivial ($\Phi_0\neq 0$, $\tilde{\Phi}_3 \neq 0$)
non-analytic in constants $F$, $G$ solutions \cite{Nambu} at $N_C
F\Lambda>2\pi^2$, $N_C G\Lambda>2\pi^2$. Thus, even at zero
current masses of fermions ($m_0=0$), as a result of the phase
transition, they acquire non-zero masses, and the massive states
of fermions correspond to the minimum of effective potential
\cite{Jona}, \cite{Coleman}. The appearance of the non-zero vacuum
condensate $\tilde{\Phi}_3 \neq 0$ indicates the spontaneous CP
symmetry breaking.

From Eqs. (12) one arrives at the relationship
\begin{equation}
Fm_0 = (F-G)f_0\Phi_0 . \label{14}
\end{equation}
It follows from Eq. (14) that if the bare masses of fermions
$m_0=0$ and $\Phi_0\neq 0$, the equality $F=G$ is valid, and we
come to the model with only one coupling constant.

The parameter cutoff ($\Lambda$) specifies the region of non-local
fermion-antifermion (quark-antiquark) interactions and this region
is determined by the size $1/\Lambda$.

Expending the logarithm in Eq. (5) in small fluctuations of fields
$\Phi_0' (x)$, $\tilde{\Phi}'_a (x)$ one obtains:
\[
\Phi_0 (x)=\Phi_0 +\Phi_0 '(x),\hspace{0.3in} \tilde{\Phi}_3
(x)=\tilde{\Phi}_3 +\tilde{\Phi}_3 '(x) ,
\]
where condensates $\Phi_0$, $\tilde{\Phi}_3$ are the solutions of
gap equations (12). Then the effective action (5) becomes
\[
S[\Phi']=-\frac{1}{2}\int d^4 xd^4 y\Phi_A
'(x)\Delta^{-1}_{AB}(x,y)\Phi_B '(y) +
\sum_{n=3}^{\infty}\frac{i}{n}\mbox{tr}\left[S_f\left(f_0\Phi_0 '+
ig_0\tilde{\Phi}_a '\gamma_5\tau^a\right)\right]^n ,
\]
\vspace{-8mm}
\begin{equation} \label{15}
\end{equation}
\vspace{-8mm}
\[
\Delta^{-1}_{AB}(p)=-ig_A g_B \mbox{tr}\left[\int\frac{d^4
k}{(2\pi^4)} S_f (k)T_A S_f (k-p)T_B\right]+\delta_{AB}M^2_A ,
\]
where $g_A=(f_0,g_0)$, $M_A=(M,\mu)$, $T_A=(1,i\gamma_5\tau^a)$,
$\Phi_A =(\Phi_0 ,\tilde{\Phi}_a)$. Using the quark propagator
(9), gap equations (12), and calculating the nonzero elements
$\Delta^{-1}_{AB}(p)$ in Eq. (15), we find

\[
\Delta_{11}^{-1}(p)=\Delta_{22}^{-1}(p)=p^2 Z_3^{-1}+{\cal
O}(g_0^2),
\]
\[
\Delta_{33}^{-1}(p)=\left(p^2+4g_0^2
\tilde{\Phi}_3^2\right)Z_3^{-1}+{\cal O}(g_0^2),
\]
\vspace{-8mm}
\begin{equation} \label{16}
\end{equation}
\vspace{-8mm}
\[
\Delta_{00}^{-1}(p)= \frac{2f_0^2 m_0 I}{m_0
-m_1}+\left(p^2+4m_1^2\right)\frac{f_0^2}{g_0^2}Z_3^{-1}+{\cal
O}(f_0^2),
\]
\[
\Delta_{03}^{-1}(p)=-4m_1 f_0 \tilde{\Phi}_3 Z_3^{-1}+{\cal O}(f_0
g_0),
\]
where the constant of renormalization is given by
\begin{equation}
Z_3^{-1} =-\frac{ig_0^2 N_C}{4\pi^4}\int \frac{d^4
q}{\left(q^2+m^2\right)^2} . \label{17}
\end{equation}
Now we introduce the renormalized fields $\tilde{\Phi}_a
(x)=Z_3^{-1/2}\tilde{\Phi}'_a(x)$, $\Phi_0
(x)=(f_0/g_0)Z_3^{-1/2}\Phi_0 '(x)$ and coupling constants
$g^2=Z_3 g_0^2$, $f^2=Z_3 f_0^2$. Logarithmic and quadratic
integrals are connected by the relation \cite{Scadron}
\begin{equation}
Z_3^{-1} =\frac{g_0^2}{m^2}I - \frac{g_0^2}{4\pi^2} . \label{18}
\end{equation}
It follows from Eq. (17) that the expansion in $g^2/4\pi^2$,
$f^2/4\pi^2$ corresponds to the $1/N_C$ expansion. Neglecting the
terms (radiation corrections) ${\cal O}(g^2)$, ${\cal O}(f^2)$,
${\cal O}(fg)$, one finds the renormalized, quadratic in the boson
fields, effective Lagrangian
\begin{equation}
{\cal L}_{free} =-\frac{1}{2}\left[\left(\partial_\mu \Phi_A
(x)\right)^2 +m^2_{AB}\Phi_A (x)\Phi_B (x)\right] , \label{19}
\end{equation}
where $\Phi_A=(\Phi_0,\tilde{\Phi}_a)$ and the elements of the
mass matrix read
\[
m_{00}^2=4m_1^2+\frac{2m_0 m^2}{m_0-m_1}
,\hspace{0.3in}m_{11}^2=m_{22}^2=0 ,
\]
\vspace{-8mm}
\begin{equation}\label{20}
\end{equation}
\vspace{-8mm}
\[
m_{03}^2=-4m_1 g\tilde{\Phi}_3
,\hspace{0.3in}m_{33}^2=4g^2\tilde{\Phi}_3^2 .
\]
The mass spectrum of the bosonic fields $\Phi_A (x)$ can be
obtained by diagonalizing the mass matrix $m_{AB}$ in Eq. (19). In
accordance with Eq. (20), the masses of the fields $\tilde{\Phi}_1
(x)$, $\tilde{\Phi}_2 (x)$ are zero, in agreement with the
Goldstone theorem \cite{Goldstone}. The fields $\Phi_0 (x)$,
$\tilde{\Phi}_3 (x)$ possess nonzero masses. One may make the
$SO(2)$-transformations
\begin{equation}
\Phi'_0 (x)=\Phi_0 (x)\cos\alpha -\tilde{\Phi}_3 (x)\sin\alpha
,\hspace{0.2in}\tilde{\Phi}'_3 (x)=\Phi_0 (x)\sin\alpha
+\tilde{\Phi}_3 (x)\cos\alpha , \label{21}
\end{equation}
where $\tan2\alpha=2m_{00}^2/(m_{33}^2-m_{00}^2)$ to diagonalize
the mass matrix $m_{AB}$. As a result, we obtain the masses of
bosonic fields $\Phi'_0 (x)$, $\tilde{\Phi}'_3 (x)$:
\[
m_{00}'^2=m_{00}^2\cos^2 \alpha+m_{33}^2 \sin^2 \alpha
-m_{03}^2\sin 2\alpha ,
\]
\vspace{-8mm}
\begin{equation}
\label{22}
\end{equation}
\vspace{-8mm}
\[
m_{33}'^2=m_{00}^2\sin^2 \alpha+m_{33}^2 \cos^2 \alpha +m_{03}^2
\sin 2\alpha ,
\]
but the fields $\tilde{\Phi}_1 (x)$, $\tilde{\Phi}_2 (x)$ remain
massless. According to Eq. (22) the field $\tilde{\Phi}_3 (x)$
acquires small mass due to the CP-violating condensate
$\tilde{\Phi}_3$. The transformations of the collective fields
(21) are induced by the corresponding transformations of fermionic
fields $\psi (x)$.

Now we evaluate the effective Lagrangian of interacting bosonic
fields by counting the components in the sum (15) with $n=3$ and
$n=4$ in the case of small parameters of expansion $g^2/4\pi^2<1$,
$f^2/4\pi^2<1$. Note that the fermion loops at $n>4$ give small
convergent corrections. We make calculations of three- and
four-point functions at the particular case of equal coupling
constants $F=G$ ($f=g$) and skipping the small CP-violating
condensate $\tilde{\Phi}_3 =0$. After renormalization, we obtain
the effective Lagrangian of interacting bosonic fields to an
accuracy ${\cal O}(g^2)$, as follows (see also \cite{Kruglov90}):
\[
{\cal L}_{int}(x) =2gm\left( \Phi_0^3 (x)+\Phi_0
(x)\tilde{\Phi}^2_a (x)\right)
\]
\vspace{-8mm}
\begin{equation}
\label{23}
\end{equation}
\vspace{-8mm}
\[
 -\frac{g^2}{2}\left[\Phi_0^4
(x)+\left(\tilde{\Phi}_a^2 (x)\right)^2  +6\Phi_0^2
(x)\tilde{\Phi}_a^2 (x)\right] .
\]

We have just considered the NJL model with $SU(2)\bigotimes U(1)$
internal symmetry group and two coupling constants which can
provide the dynamical CP symmetry violation. The scalar collective
field $\Phi_0$ is associated with $\sigma$-meson, and the
pseudoscalar bosons $\tilde{\Phi}_a$ are associated with the
triplet of pions. As usual \cite{Nambu}, pions play the role of
the Goldstone bosons. The four-fermion interaction is some
approximation to the real quark interactions. However, to take
into consideration confinement of quarks, one has to modify the
model.

\end{document}